\documentclass[aps,prl,twocolumn,amsmath,showpacs]{revtex4}
\usepackage{graphicx}
\begin{document}

\title{Nonequilibrium phase transition on a randomly diluted lattice}

\author{Thomas Vojta}
\author{Man Young Lee}
\affiliation{Department of Physics, University of Missouri-Rolla, Rolla, MO 65409}

\date{\today}

\begin{abstract}
We show that the interplay between geometric criticality and dynamical fluctuations leads
to a novel universality class of the contact process on a randomly diluted lattice. The
nonequilibrium phase transition across the percolation threshold of the lattice is
characterized by unconventional activated (exponential) dynamical scaling and strong
Griffiths effects. We calculate the critical behavior in two and three space dimensions,
and we also relate our results to the recently found infinite-randomness fixed point in
the disordered one-dimensional contact process.

\end{abstract}


\maketitle


Nonequilibrium systems can undergo continuous phase transitions between different steady
states. These transitions are characterized by collective fluctuations over large
distances and long times similar to the behavior of equilibrium critical points. Examples
can be found in population dynamics and epidemics, chemical reactions, growing surfaces,
and in granular flow and traffic jams (for recent reviews see, e.g., Refs.\
\cite{SchmittmannZia95,Marro_book,Dickman,Hinrichsen00,Odor,tauber_rev}).

If a nonequilibrium process is defined on a randomly diluted spatial lattice, its
dynamical fluctuations coexist with geometric fluctuations. Site or bond dilution defines
a percolation problem for the lattice with a geometric phase transition at the
percolation threshold \cite{percolation}. In this Letter we address the question of how
the interplay between geometric criticality due to percolation and dynamical fluctuations
of the nonequilibrium process influences the properties of the phase transition.

Our starting point is the contact process \cite{contact}, a prototypical system
exhibiting a nonequilibrium phase transition. It is defined on a $d$-dimensional
hypercubic lattice ($d\ge 2$). Each site  can be active (occupied by a particle) or
inactive (empty). In the course of the time evolution, active sites  infect their
neighbors, or they spontaneously become inactive. Specifically, the dynamics is given by
a continuous-time Markov process during which particles are created at empty sites at a
rate $\lambda n/ (2d)$ where $n$ is the number of active nearest neighbor sites.
Particles are annihilated at unit rate.
For small birth rate $\lambda$, annihilation dominates, and the absorbing state without
any particles is the only steady state (inactive phase). For large birth rate $\lambda$,
there is a steady state with finite particle density (active phase).  The two phases are
separated by a nonequilibrium phase transition in the directed percolation
\cite{dp,conjecture} universality class at some $\lambda=\lambda_c^0$.

\begin{figure}[b]
\includegraphics[width=7.3cm]{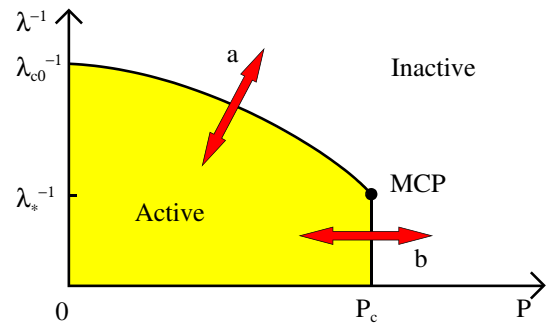}
\caption{Schematic phase diagram of a site diluted contact process
 as function of impurity concentration $p$ and birth rate $\lambda$. There is a multicritical point at $p=p_c$ and
 $\lambda=\lambda_*$. The phase transition (b) across the percolation threshold of the lattice is the topic of this Letter.}
\label{fig:pd}
\end{figure}

We introduce quenched site dilution \cite{GEP} by randomly removing lattice sites with
probability $p$. The resulting phase diagram of the site-diluted contact process is
sketched in Fig.\ \ref{fig:pd}.
For small impurity concentrations below the percolation threshold of the lattice,
$p<p_c$, the active phase survives, but the critical birth rate increases with $p$ (to
compensate for the missing neighbors). Right at the percolation threshold the active
phase survives on the infinite percolation cluster for $\lambda>\lambda_*$. The (multi)
critical birthrate $\lambda_*$ must be smaller than the critical birthrate of the
one-dimensional (1D) contact process because the critical percolation cluster is
connected, infinitely extended, and its fractal dimension is $D_f>1$. For $p>p_c$, no
active phase can exist because the lattice consists of disconnected clusters of finite
size that do not support a steady state density of active sites.

The contact process on a site-diluted lattice therefore has two nonequilibrium phase
transitions, separated by a multicritical point. For $p<p_c$, the transition (marked by
``a'' in Fig.\ \ref{fig:pd}) is expected to be in the universality class of the generic
disordered contact process \cite{Noest86,janssen97,moreira} which has reattracted
considerable attention recently \cite{hooyberghs,vojtadickison05}. In contrast, the phase
transition across the percolation threshold $p_c$ of the lattice for $\lambda>\lambda_*$
(transition ``b'' in Fig.\ \ref{fig:pd}) has received much less attention.

In this Letter, we show that the interplay between geometric criticality and dynamic
fluctuations leads to a novel universality class for this nonequilibrium phase
transition. Even though the transition is driven entirely by the geometry of the lattice,
the dynamical fluctuations of the contact process enhance the singularities in all
quantities involving dynamic correlations. Our results can be summarized as follows. The
dynamical scaling is not of conventional power-law form but activated, i.e., the relation
between correlation length $\xi_\perp$ and correlation time $\xi_\parallel$ is
exponential,
\begin{equation}
\ln \xi_\parallel \sim \xi_\perp^\psi~ \label{eq:activated}
\end{equation}
with the critical exponent $\psi$ being equal to the fractal dimension of the critical
percolation cluster, $\psi =D_f$. As a result, the long-time decay of the density $\rho$
of active sites at $p=p_c$ is ultra-slow,
\begin{equation}
\rho(t)  \sim [\ln(t/t_0)]^{-\bar\delta}~.
\label{eq:rhot}
\end{equation}
The exponent $\bar\delta=\beta_c/(\nu_c D_f)$ is determined by $D_f$ together with the
order parameter and correlation length exponents, $\beta_c$ and $\nu_c$, of the lattice
percolation transition \cite{subscript}. In contrast to the enhanced dynamical
singularities, the exponents of static quantities like the steady state density
$\rho_{st}$  and the spatial correlation length $\xi_\perp$ are identical the
corresponding lattice percolation exponents,
\begin{eqnarray}
\rho_{st} (p) &\sim& |\Delta|^{\beta_c}  \qquad (\Delta<0)~, \label{eq: rhost}\\
\xi_\perp &\sim& |\Delta|^{\nu_c}~\label{eq:xiperp}.
\end{eqnarray}
where $\Delta=p-p_c$ measures the distance from the percolation threshold.
Off criticality, i.e., away from the percolation threshold, we find strong
Griffiths effects \cite{Griffiths} characterized by a non-exponential density decay \cite{Noest86},
\begin{eqnarray}
\rho(t) \qquad~ &\sim& (t/t_0)^{-d/z'} ~\qquad\qquad (p>p_c) \label{eq:griffiths+}\\
\rho(t)-\rho_{st} &\sim& e^{ -\left[(d/z'') \ln(t/t_0) \right]^{1-1/d}}  \qquad (p<p_c)~,
\label{eq:griffiths-}
\end{eqnarray}
where the nonuniversal exponents $z'$ and $z''$ diverge as $z', z'' \sim \xi_\perp^{D_f}$
for $p \to p_c$.
In the remainder of the Letter we sketch the derivation of these results and calculate
exponent values and additional observables. We also relate our results to the disordered
1D contact process \cite{hooyberghs,vojtadickison05} and to the diluted quantum Ising
model \cite{Senthil}.

Let us start by considering the steady state density $\rho_{st}$ of active sites (i.e.,
the order parameter of the transition). A nonzero steady state density can only develop
on the infinite percolation cluster; finite clusters do not contribute because they
eventually go into the inactive state via a rare fluctuation. For $\lambda>\lambda_c$,
the infinite cluster is in the active phase. The total steady state density is
proportional to the number of sites in the infinite cluster, $\rho_{st} \sim P_{\infty
}(p)\sim (p_{c}-p)^{\beta _{c}}$. The order parameter exponent of the nonequilibrium
transition is therefore identical to that of the lattice percolation transition, $\beta
=\beta _{c}$, as stated in (\ref{eq: rhost}). To determine the spatial correlation length
$\xi_\perp $ we note that the correlations of the contact process cannot extend beyond
the connectedness length $\xi _{c}$ of the percolating lattice because sites on different
percolation clusters are decoupled. On the other hand, for $\lambda>\lambda_*$, all sites
on the same cluster are strongly correlated in space even though they collectively
fluctuate in time. We thus conclude $\xi \sim \xi _{c}$ and $\nu =\nu _{c}$ in agreement
with (\ref{eq:xiperp}).

We now study the time dependence of the density $\rho(t)$ of active sites, starting from
a completely active lattice.  We first consider the contact process on a single
percolation cluster of finite size (number of sites) $s$. For $\lambda>\lambda_*$ such a
cluster is locally in the active phase. It therefore has a metastable state with a
nonzero density of active sites. This metastable state can decay into the inactive state
only via a rare collective fluctuation involving all sites of the cluster. The
probability for such a rare event decreases exponentially with the size $s$ of the
cluster. Therefore, the life time $t_s$ of the metastable active state on a cluster
increases exponentially with its size $s$,
\begin{equation}
 t_s(s) \sim t_0 e^{A(\lambda) s} \label{eq:ts}~,
\end{equation}
where $t_0$ is some microscopic time scale. The prefactor $A(\lambda)$ vanishes at the
multicritical point, $A(\lambda_*)=0$, and increases with increasing $\lambda$. The
number of sites $s$ of a percolation cluster is connected to its linear size $R_s$ via
$s\sim R_s^{D_f}$. Therefore, (\ref{eq:ts}) establishes the exponential relation between
length and time scales, $\ln t_s \sim R_s^{D_f}$ leading to activated dynamical scaling
(\ref{eq:activated}).

After having analyzed a single cluster we turn to the full percolation problem. From
classical percolation theory, we know that close to $p_{c}$ the number $n_{s}$ of
occupied clusters of size $s$ per lattice site (excluding the infinite cluster for
$p<p_c$) obeys the scaling form
\begin{equation}
n_{s}\left( \Delta\right) =s^{-\tau_c }f\left( s\Delta^{1/\sigma_c }\right) .
\label{eq:percscaling}
\end{equation}
The scaling function $f(x)$ behaves as
\begin{eqnarray}
f(x) &\sim& \exp({-B_1 x}) \qquad\qquad (p>p_c)\\ f(x) &=& {\rm const} \qquad\qquad
(p=p_c)\\ f(x) &\sim& \exp[{-(B_2 x)^{1-1/d}}] \qquad (p<p_c).
\end{eqnarray}
where $B_1,B_2$ are constants. The exponents $\tau_c $ and $\sigma_c$ determine all
critical exponents of the percolation transition of the lattice including the correlation
length exponent $\nu_c =({\tau_c -1})/{(d\sigma_c )}$, the order parameter exponent
$\beta_c=(\tau_c-2)/\sigma_c$, and the fractal dimension $D_f=d/(\tau_c-1)$ of the
percolating cluster \cite{percolation}.

In order to obtain the total density of active sites for the contact process on the diluted lattice,
we sum the number of active sites
over all percolation clusters. Combining the cluster size distribution
(\ref{eq:percscaling}) with the lifetime of the metastable active state (\ref{eq:ts})
leads to
\begin{eqnarray}
\rho(t,\Delta) &\sim& \int ds ~ s ~ n_s(\Delta) \exp[-t/t_s(s)]
\label{eq:rhotdelta}
\end{eqnarray}
Right at the percolation threshold, this reduces to
\begin{equation}
\rho(t,0) \sim \int ds ~s^{1-\tau_c} \exp[-(t/t_0e^{As})]~.
\end{equation}
The leading behavior of this integral can be found by noticing that
only islands with size $s>s_{\rm min}(t)= A^{-1} \ln(t/t_0)$ contribute at
time $t$. The critical long-time dependence of the total density is thus
given by
\begin{equation}
\rho(t,0) \sim [\ln(t/t_0)]^{2-\tau_c} \qquad (p=p_c).
\end{equation}
This completes the derivation of (\ref{eq:rhot}) with the critical exponent $\bar\delta$
given by $\bar\delta=\tau_c-2 = \beta_c/(\nu_c D_f)$ in agreement with general scaling
arguments \cite{hooyberghs,vojtadickison05}.

We now consider the behavior of the density off criticality. In the inactive phase,
$p>p_c$, the time dependence of the density is given by
\begin{equation}
\rho(t,\Delta) \sim  \int ds s^{1-\tau_c} \exp [-B_1 s \Delta^{1/\sigma_c} -(t/t_0 e^{A
s})].
\end{equation}
For long times, the leading behavior of the integral can be calculated using the
saddle-point method, giving
\begin{equation}
\rho(t,\Delta) \sim t^{-(B_1/A) \Delta^{1/\sigma_c}} \qquad (p>p_c)
\end{equation}
equivalent to  (\ref{eq:griffiths+}). The nonuniversal exponent $z'$ is given by  $z' =(
Ad/B_1)\Delta^{-1/\sigma_c} \sim \xi_\perp^{D_f}$.

In the active phase, $p<p_c$, there is a nonzero steady state density $\rho_{st}$ coming
from the infinite percolation cluster. However, the approach of the density towards this
value is still determined by the slow decay of the metastable states of the finite
percolation clusters
\begin{eqnarray}
&&\rho(t,\Delta) -\rho_{st}(\Delta) \sim
\\
&\sim& \int ds s^{1-\tau_c} \exp\left[-\left(B_2 s |\Delta|^{1/\sigma_c}\right)^{1-1/d} -
(t/t_0 e^{As})\right] \nonumber
\end{eqnarray}
Using the saddle point method to calculate the leading long-time behavior gives
(for $p<p_c$)
\begin{equation}
\rho(t,\Delta) -\rho_{st}(\Delta) \sim e^{ - \left[ (B_2/A) |\Delta|^{1/\sigma_c}
\ln(t/t_0) \right]^{1-1/d} } ~.
\end{equation}
This completes the derivation of (\ref{eq:griffiths-}) with $z'' = (A d/B_2)
|\Delta|^{-1/\sigma_c} \sim \xi_\perp^{D_f}$. The non-exponential off-critical relaxation
of the density (\ref{eq:griffiths+},\ref{eq:griffiths-}) is characteristic of a Griffiths
region in the contact process \cite{Noest86,Griffiths}. We also point out that time and
spatial correlation length enter these equations in the form of the combination
$\ln(t)/\xi_\perp^{D_f}$ again characteristic of activated scaling.

We now turn to the influence of an external source field $h$ that describes spontaneous
particle creation at a rate $h$ at each lattice site. To determine the steady state
density as a function of $h$ we again start by considering a single percolation cluster
of size $s$. For $\lambda>\lambda_*$, the cluster is active if at least one particle has
been spontaneously created on one of the $s$ sites within the life time $t_s(s)=t_0
e^{As}$. For small $h$, the average number of particles created on a cluster of size $s$
within time $t_s$ is $M_s(h) =h s t_s=h s t_0 e^{As}$. If $M_s>1$, the cluster is
(almost) always active. If $M_s<1$, it is active with a probability proportional to
$M_s$.
The total steady state density is obtained by summing over all clusters
\begin{equation}
\rho_{st}(h,\Delta) \sim \int ds ~ s ~ n_s(\Delta) \min[1,M_s(h)]~.
\label{eq:rhohdelta}
\end{equation}
Evaluating this integral analogously to the  time-dependent density  (\ref{eq:rhotdelta})
yields, for small fields $h$,
\begin{eqnarray}
\rho_{st} (h)  &\sim& [\ln(h_0/h)]^{-\bar\delta}~ \qquad\qquad (p=p_c), \label{eq:rhoh}\\
\rho_{st} (h) &\sim& (h/h_0)^{d/z'} ~\qquad\qquad (p>p_c),\\
\rho_{st} (h) &\sim& \exp \left[ -(d/z'') \ln(h_0/h) \right]^{1-1/d}  \quad (p<p_c)~,
\end{eqnarray}
where $h_0 \sim 1/t_0$. At $p=p_c$, the relation between density and field is
logarithmic, as expected from activated scaling. Off criticality, we find strong
Griffiths effects similar to those in the time-dependence of the density.

The above results can also be derived from a scaling theory. In the active phase, the
density is proportional to the number of sites in the infinite percolation cluster. Thus,
its scale dimension must be $\beta_c/\nu_c$. Time must enter via the scaling combination
$\ln(t) b^{D_f}$ reflecting the exponential dependence of the life time (\ref{eq:ts}) on
the cluster size. The field $h$, being a rate, scales like inverse time. We therefore
obtain the following scaling form:
\begin{equation}
\rho[\Delta,\ln(t),\ln(1/h)] =b^{\beta_c/\nu_c}\rho[\Delta b^{-1/\nu_c}, \ln(t) b^\psi,
\ln(1/h) b^\psi]~
\end{equation}
where $b$ is an arbitrary (length) scale factor and $\psi=D_f$. This form is consistent
with all our explicit results.

All critical exponents of the nonequilibrium phase transition are determined by the
classical percolation exponents of the lattice. In two space dimensions, their values are
known exactly and in three dimensions they are known numerically with high accuracy
\cite{percolation}. Table I shows numerical exponent values for these cases.
\begin{table}[tbp]
\begin{tabular}{l|c|c}
\hline
Exponent    & $~d=2~$ & $~d=3~$ \\
\hline
$\beta  =\beta_c$                   & 5/36 & 0.417  \\
$\nu    =\nu_c$                     &  4/3 & 0.875  \\
$\psi   =D_f=d-\beta_c/\nu_c$       & 91/48& 2.523  \\
$\bar\delta=\beta_c/(\nu_c D_f) $   & 5/91 & 0.188  \\ \hline
\end{tabular}%
\caption{Critical exponents of the nonequilibrium phase transition at $p=p_c$ in two and
three space dimensions. }
\end{table}

We also briefly comment on the early time behavior. For $\lambda>\lambda_*$ each
percolation cluster is locally in the active phase. Starting from a single active seed,
the cloud of active sites thus initially grows ballistically, i.e., the radius of the
cloud grows linearly with time, until a metastable state is reached in which it covers
the entire percolation cluster. The time required for this initial spreading on an island
of size $s$ is $t_i(s) \sim R_s \sim s^{1/D_f}$. As discussed above, the metastable state
decays only at the much larger time scale $t_s(s) \sim e^{As}$. We thus arrive at the
somewhat surprising conclusion that the early time behavior of the contact process on our
diluted lattice is much faster than the logarithmically slow long-time decay of the
density.

In the remaining paragraphs, we discuss the generality of our results, compare them to
the transition in the diluted quantum Ising model \cite{Senthil} and to the recently
found infinite-randomness critical point in a random 1D contact process
\cite{hooyberghs,vojtadickison05}. We also compare to a general classification of phase
transitions with quenched disorder \cite{VojtaSchmalian05}.

The logarithmic time and field dependencies (\ref{eq:rhot}) and (\ref{eq:rhoh}) at the
nonequilibrium phase transition at $p=p_c$ as well as the strong Griffiths effects in its
vicinity are the direct result of combining the spectrum of percolation cluster sizes
(\ref{eq:percscaling}) with the exponential dependence (\ref{eq:ts}) of the life time on
the cluster size. We therefore expect similar behavior in other diluted equilibrium or
nonequilibrium systems that share this exponential relation between length and time
scales.  One example is the diluted transverse field Ising model \cite{Senthil}. In this
system, the energy gap of a cluster decreases exponentially with its size. As a result,
the scaling behavior at the quantum phase transition across the percolation threshold of
the lattice is very similar to the one found in this paper.

Recently, the critical point of the 1D contact process with spatial disorder was found to
be of infinite-randomness type \cite{hooyberghs,vojtadickison05}. Analogous behavior is
expected for the generic disordered directed percolation transition in higher dimensions,
e.g., the transition at $p<p_c$ in the site-diluted contact process (transition a in
Fig.\ \ref{fig:pd}). Our critical point at the percolation threshold shares some
characteristics with these infinite-randomness critical points, notably the exponential
relation between correlation length and time as well as the logarithmically slow decay of
the total density. However, it belongs to a different universality class with novel
critical exponents. Moreover, the early time behavior is different (logarithmically slow
at the generic infinite randomness critical point but of power-law type at our
transition).

Lastly, we point out that our results are in agreement with a general classification of
phase transitions with quenched disorder (and short-range interactions) according to the
effective dimensionality $d_{eff}$ of the droplets or clusters \cite{VojtaSchmalian05}.
Three cases can be distinguished: (i) If the clusters are below the lower critical
dimension of the problem, $d_{eff}<d_c^-$, the critical behavior is of conventional
power-law type and the Griffiths effects are exponentially weak. (ii) If $d_{eff}=d_c^-$,
the critical point shows activated scaling accompanied by strong, power-law Griffiths
effects. This case is realized in random transverse field Ising magnets
\cite{Senthil,dsf9295} as well as in our diluted contact process. (iii) If
$d_{eff}>d_c^-$, the phase transition is smeared because locally ordered clusters can
undergo the phase transition independently from the bulk. This occurs, e.g., for some
metallic quantum magnets \cite{us_rounding} or for the contact process with extended
defects \cite{contact_pre,Dickison05}.

In conclusion, we have shown that the contact process on a diluted lattice has unusual
properties. The interplay between geometric criticality and dynamical fluctuations leads
to a novel universality class with activated scaling and ultraslow dynamics.
Interestingly, despite its ubiquity in theory, experimental observations of directed
percolation scaling \cite{hinrichsen_exp} are very rare. Our results suggest that
peculiar disorder effects may be responsible for this in at least some of the
experiments.

This work has been supported in part by the NSF under grant nos. DMR-0339147 and
PHY99-07949, by Research Corporation and by the University of Missouri Research Board.
Parts of this work have been performed at the Aspen Center for Physics and the Kavli
Institute for Theoretical Physics, Santa Barbara.


\begin{thebibliography}{99}
\bibitem{GEP} Our percolating lattice is externally given.
    This differs from the general epidemic process with
    immunization where the nonequilibrium process itself dynamically generates
    a percolation cluster of inaccesible sites \cite{GEP1}.
\bibitem{subscript} In the following, lattice percolation exponents carry a subscript
    $c$ while the exponents of the nonequilibrium transition of the contact process do not
    carry a subscript.
\end{thebibliography}
\end{document}